\documentclass[showpacs,amssymb,floatfix,twocolumn]{revtex4}[14pt]
\usepackage{graphicx}
\usepackage{dcolumn}
\usepackage{graphics}
\usepackage{epstopdf}
\usepackage{pdfpages}
\usepackage{amssymb}
\usepackage{amsmath}
\usepackage{bm}

\begin{document}
\title{Evolution of the resonances of two parallel dielectric cylinders with distance between them}
\author{E.N. Bulgakov, K.N. Pichugin, A.F. Sadreev}
\affiliation{Kirensky Institute of Physics, Federal Research Center KSC Siberian Branch,
Russian Academy of Sciences,  Krasnoyarsk 660036, Russia}
\date{\today}
\begin{abstract}
We study behavior of resonant modes under change of the distance between two parallel dielectric
cylinders. The processes of mutual scattering of Mie resonant modes by cylinders result in
an interaction between the cylinders which
lifts a degeneracy of resonances of the isolated cylinders.
There are two basic scenarios of evolution of resonances with the distance.
For strong interaction  of cylinders resonances  are unbound with
avoided crossings from the of the Mie resonances with increasing of the distance.
That scenario is typical for low exciting resonances (monopole and dipole).
For weak interaction of cylinders the resonances are bound around
highly excited Mie resonances. Both scenarios
demonstrate a significant enhancement of the $Q$ factor compared to the case of isolated cylinder.

\end{abstract}
 \maketitle
\section{Introduction}
It is rather challenging for optical resonators to support
resonances of simultaneous subwavelength mode volumes and high
$Q$ factors. The traditional way for increasing the $Q$ factor of
optical cavities is a suppression of leakage of resonance mode
into the radiation continua.   That is achieved usually by
decreasing the coupling of the resonant mode with the continua
by the use of metals, photonic band gap structures, or
whispering-gallery-mode resonators. All of these approaches lead
to reduced device efficiencies because of complex designs,
inevitable metallic losses, or large cavity sizes. On the
contrary,   all-dielectric   subwavelength nanoparticles  have
 recently   been   suggested   as   an   important   pathway   to
enhance capabilities of traditional nanoscale resonators by exploiting
the multipolar Mie resonances being  limited
only by radiation losses \cite{Kuznetsov2016,Koshelev2018}.

The decisive breakthrough came with the paper by Friedrich and
Wintgen \cite{Friedrich1985} which put forward the idea of
destructive interference of two neighboring resonant modes leaking
into the continuum. Based on a simple generic two-level model they
formulated the condition  for the bound state in the continuum
(BIC) as the state with zero resonant width for crossing of
eigenlevels of the cavity or avoided crossing of resonances.
This principle was later explored in
open plane wave resonator where the BIC occurs in the vicinity of
degeneracy of the closed integrable resonator \cite{SBR}.

However, these BICs exist provided that they embedded into a
single continuum of propagating modes of a directional waveguide.
In photonics the optical BICs embedded into the radiation
continuum can be realized by two ways. The first way is realized
in an optical cavity coupled with the continuum of 2d photonic
crystal (PhC) waveguide \cite{BS2008} that is an optical variant
of microwave system \cite{SBR}. Alternative way is the use
periodic PhC systems (gratings) or arrays of dielectric particles
in which resonant modes leak into a restricted number of
diffraction continua \cite{Shipman2005,Marinica,Hsu2013,PRA2014,PRA96}. Although the
exact BICs can exist only in infinite periodical arrays
\cite{Colton,Silveirinha14}, finite arrays demonstrate
resonant modes with the very high $Q$ factor which grows
quadratically \cite{Sadrieva2019} or even cubically \cite{Bulgakov2019a}
with the number of particles (quasi-BICs).

Isolated subwavelength high-index dielectric resonators are more advantageous
from an applied point of view to achieve high $Q$ resonant modes (super cavity modes)
\cite{Rybin2017,Koshelev2018,Bogdanov2019}.  Such super cavity modes originate
from avoided crossing of the resonant modes, specifically
the Mie-type resonant mode and the Fabry-P\'{e}rot
resonant mode under variation of the aspect ratio of the
dielectric disk which could result in a significant enhancement of the $Q$ factor.
It is worthy also to
notice  the idea of formation of long-lived, scar like modes near
avoided resonance crossings in optical deformed microcavities
\cite{Wiersig2006}. The dramatic  $Q$ factor enhancement was
predicted by Boriskina \cite{Boriskina2006,Boriskina2007} for
avoided crossing of very highly excited whispering gallery modes
in symmetrical photonic molecules of dielectric disks on a
surface.

In the present paper we consider a similar way to enhance the
$Q$ factor by variation of the distance between two identical
dielectric cylinders parallel each other as sketched in Fig. \ref{fig2}.
 As different from papers
\cite{Wiersig2006,Boriskina2006,Boriskina2007,Unter2008,Benyoucef2011}
we consider the avoided crossing of
low excited resonant modes (monopole, dipole and quadruple) with variation of
the distance between two cylinders. Because of lifting of the axial symmetry
many Mie resonances contribute into the resonances of two cylinders which
show two basic scenarios of evolution  with the distance, bound to the Mie resonances and unbound.

\section{Avoided crossing under variation of distance between two cylinders}

\begin{figure}
\includegraphics*[width=7cm,clip=]{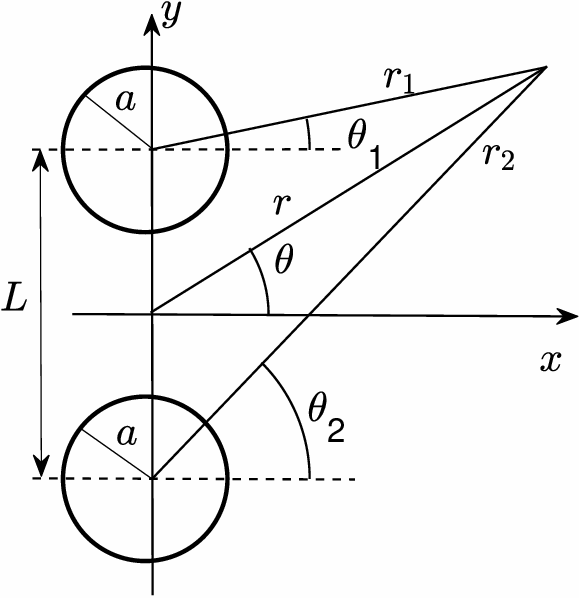}
\caption{Two identical parallel dielectric cylinders with radii $a$ and refractive
index $\sqrt{\epsilon}=\sqrt{30}$.}
\label{fig2}
\end{figure}

The problem of scattering of electromagnetic waves from two parallel infinitely long
dielectric cylinders sketched in Figs. \ref{fig2} was solved
long time ago \cite{Olaofe1970,Young1975,Tsuei1988}.
The solutions for electromagnetic field (the component of electric field directed
along the cylinders $\psi=E_z$) (see Fig. \ref{fig2})
outside the cylinders are given by
\begin{eqnarray}
\label{scat}
&\psi_1=\sum_nA_{1n}H_n^{(1)}(kr_1)e^{in\theta_1},&\\
&\psi_2=\sum_nA_{2n}H_n^{(1)}(kr_2)e^{in\theta_2}.&
\end{eqnarray}
Inside the cylinders we have
\begin{eqnarray}
\label{inside}
&\psi_1=\sum_nB_{1n}J_n(\sqrt{\epsilon}kr_1)e^{in\theta_1},&\\
&\psi_2=\sum_n\sum_nB_{2n}J_n(\sqrt{\epsilon}kr_2)e^{in\theta_2}.&
\end{eqnarray}
By means of the Graf formula \cite{Yasumoto}
\begin{eqnarray}\label{Graf}
&H_n^{(1)}(kr_1)e^{in\theta_1}=\sum_mi^{n-m}H_{m-n}^{(1)}(kL)J_m(kr_2)e^{im\theta_2},&\\
&H_n^{(1)}(kr_2)e^{in\theta_2}=\sum_mi^{n-m}H_{m-n}^{(1)}(kL)J_m(kr_1)e^{im\theta_1},&
\end{eqnarray}
the total field $\psi=\psi_{inc}+\psi_1+\psi_2$
can be written completely in either coordinate system.

Applying the boundary conditions at $r_j=a$ leads to \cite{Olaofe1970,Tsuei1988}
\begin{eqnarray}
\label{AB}
&A_{1n}=i^nS_n(k)\sum_mi^{-m}H_{n+m}(kL)A_{2n},&\nonumber\\
&A_{2n}=i^nS_n(k)\sum_mi^{-m}H_{n+m}(kL)A_{1n},&
\end{eqnarray}
where $S_n$ are the scattering matrix amplitudes for the isolated cylinder
\begin{equation}
\label{Smatrix}
S_n(k)=\frac{\sqrt{\epsilon}J^{'}_{m}(\sqrt{\epsilon}ka)J_m(k)-J^{'}_{m}(k)J_m(\sqrt{\epsilon}k)}
{H^{(1)'}_{m}(k)J_m(\sqrt{\epsilon}k)-\sqrt{\epsilon}J^{'}_{m}(\sqrt{\epsilon}k)H^{(1)}_m(k)}.
\end{equation}
The resonances are given by the complex roots of the following  equation
\begin{equation}\label{eqforpole}
    Det[\hat{M}^2-I]=0
\end{equation}
where matrix elements $\hat{M}$ is given by Eq. (\ref{AB})
and equal
\begin{equation}\label{M}
    M_{mn}=S_m(k)i^{m-n}H_{m+n}(kL)
\end{equation}
and $I$ is the unit matrix.

\begin{figure}
\includegraphics*[width=10cm,clip=]{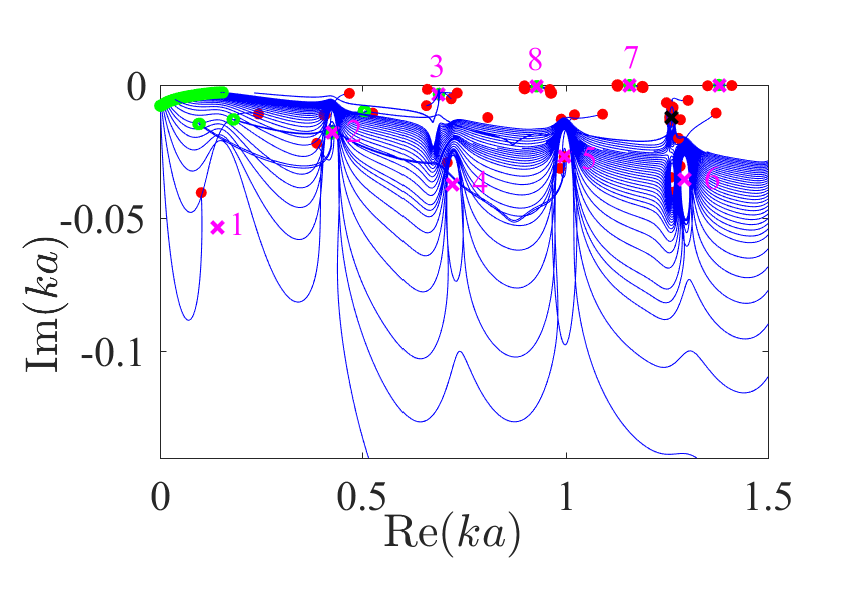}
\caption{The dependence of resonant frequencies (poles) on the
distance between cylinders.  Open green circles correspond $L=1000a$,
closed red circles correspond to minimal distance $L=2a$, where $a$ is the radius of
cylinders and black crosses are the enumerated resonant frequencies of the isolated cylinder
taken from Fig. \ref{fig3}.}
\label{fig4}
\end{figure}

\begin{figure}
\includegraphics[width=9cm,clip=]{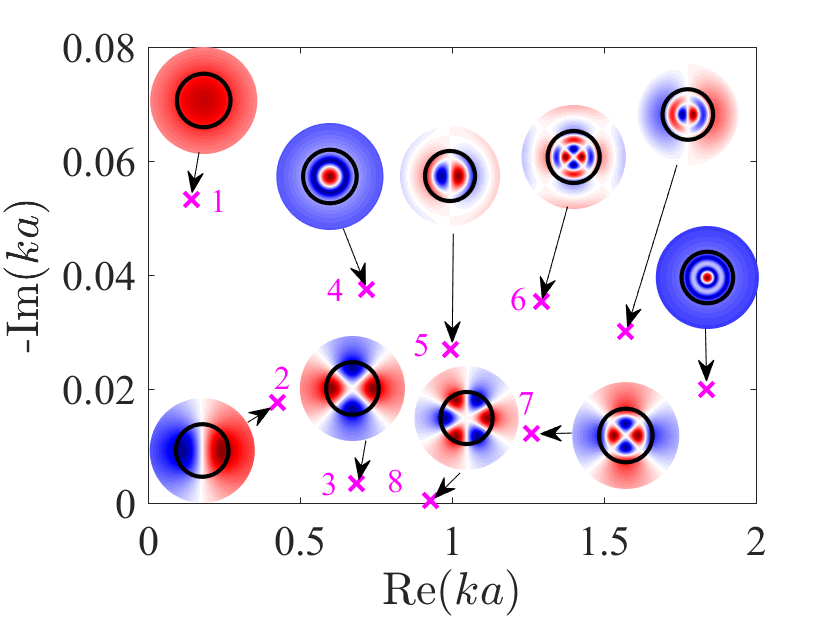}
\caption{The dimensionless Mie resonances (close circles) and corresponding
resonant modes (the component $E_z$) of isolated cylinder.}
\label{fig3}
\end{figure}

In general Eq. (\ref{eqforpole}) has an infinite number of complex resonant frequencies (poles)
$k=k_n-i\gamma_n$ which are shown in Fig. \ref{fig4}.
First of all one can see that major part of resonances evolves with the distance bypassing
the Mie resonances  of the isolated cylinder marked by crosses some part of which are collected
in Fig. \ref{fig3}.
Second, there are a small part of resonances  which are bound around the Mie resonances
with small imaginary parts. As a rule these Mie resonances correspond to highly excited Mie resonances.

\begin{figure}
\includegraphics[width=5.5cm,clip=]{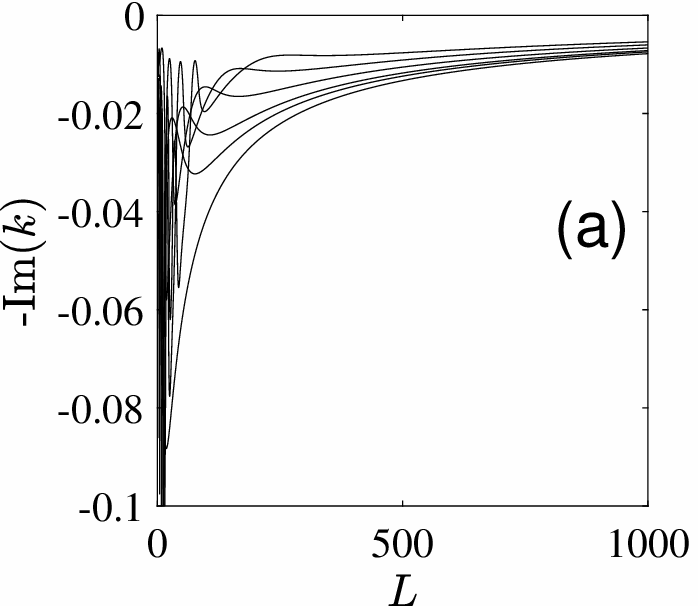}
\includegraphics[width=5cm,clip=]{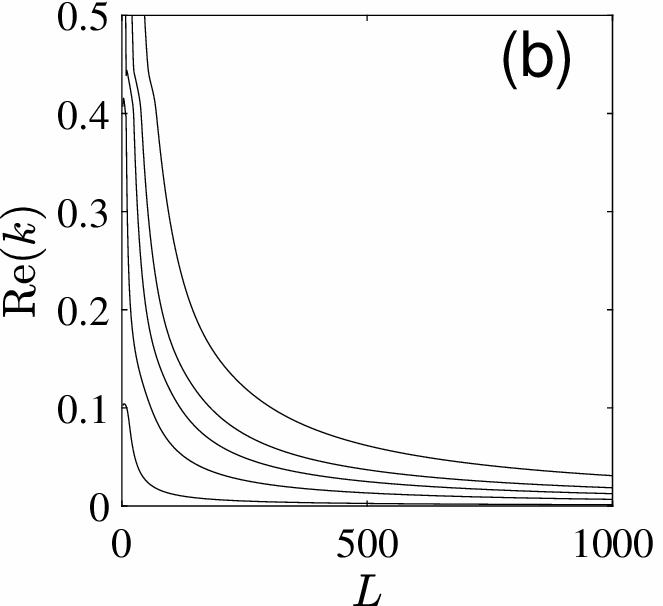}
\caption{The behavior of resonant widths and positions of a few lowest
resonances shown in Fig. \ref{fig4} on the distance between cylinders.}
\label{ImRe}
\end{figure}

Let us consider the asymptotic behavior of poles for $L\rightarrow \infty$.
By use of asymptotical behavior of the Hankel functions \cite{Abramowitz1964}
we have for matrix (\ref{M}) the following
\begin{equation}\label{Maympt}
    M_{mn}\sim \sqrt{\frac{2}{\pi kL}}e^{i(kL-\pi/4)}S_m(k)(-1)^n .
\end{equation}
Let us take the eigenvector of  matrix $\widehat{M}$ $\overrightarrow{\psi}^{+}=
(\psi_1, \psi_2, \psi_3,\ldots)$. Then
Eq. (\ref{eqforpole}) takes the following form
\begin{equation}\label{poleasympt}
    \sqrt{\frac{2}{\pi kL}}e^{ikL}S_m(k)\sum_n(-1)^n\psi_n=\pm \psi_m,
\end{equation}
which has the solution provided that
\begin{equation}\label{poleasympt1}
    \sqrt{\frac{2}{\pi kL}}e^{ikL}\sum_n(-1)^nS_n(k)=\pm 1.
\end{equation}
For the absolute value we have
\begin{equation}\label{poleabs}
    \frac{2e^{\gamma_n L}}{\pi |k_n|L}=\frac{1}{|\sum_m(-1)^mS_m(k_n)|^2}
\end{equation}
where $\gamma_n=-2{\rm Im}(k_n)$.
The poles of the isolated cylinder are given by poles of the S-matrix, i.e., by
equation $\frac{1}{S_m(k)}=0$.
Therefore from (\ref{poleabs}) it follows that, first, the poles of double cylinders do not converge
to the poles of the isolated cylinder. Second, the line widths and resonant positions
given by imaginary parts and real parts of the complex
resonant frequencies respectively are to limit to zero as indeed Fig. \ref{ImRe} (a) and (b) illustrates.
Therefore we can conclude that the resonances of two dielectric
cylinders do NOT limit to the Mie resonances of the isolated cylinders at $L\rightarrow\infty$.
The reason is related to the exponential factor $\exp(\gamma_nL)$ of the resonant modes
(the Gamov states).

Next, we consider behavior of some typical resonances in Fig. \ref{fig4} in detail
from the limiting case $L=2a$ to $L=1000a$.
Due to the symmetry relative to $x\rightarrow -x$ and $y\rightarrow -y$
the resonant modes can be classified as $\psi_{\sigma,\sigma'}$ where the
indices $\sigma=s, a$ respond for symmetric and antisymmetric  modes respectively.
We start with the first two lowest symmetric and antisymmetric monopole resonant modes
$\psi_{s;s/a}$.
\begin{figure}
\includegraphics[width=7cm,clip=]{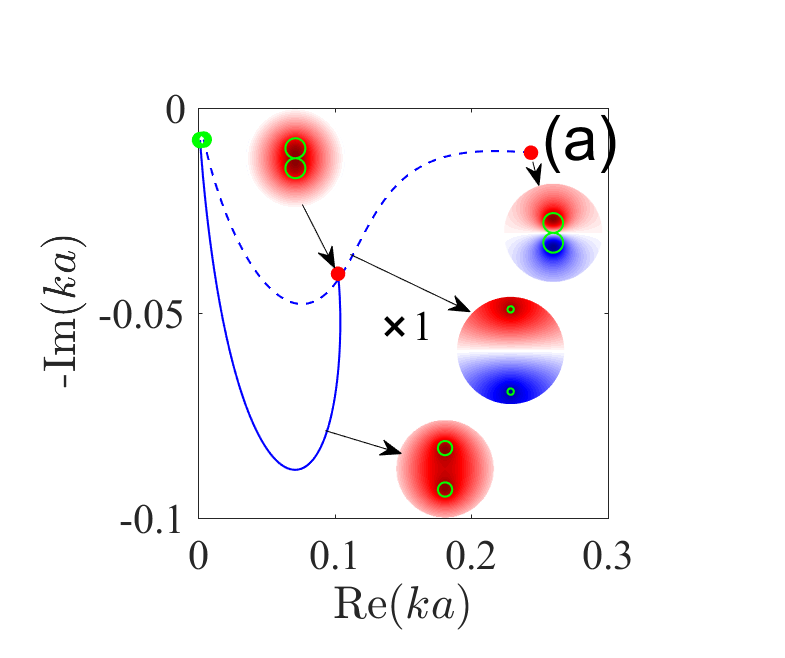}
\includegraphics[width=5.3cm,clip=]{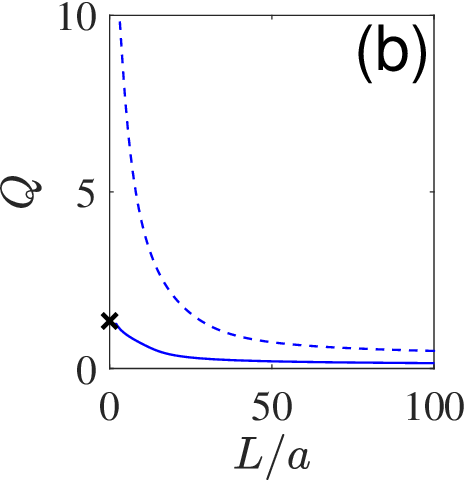}
\caption{Evolution of resonant frequencies and monopole modes $\psi_{s,s/a}^{(0)}$ (a)
and the $Q$ factors with the  distance between the cylinders. Solid/dash line show
symmetric/antisymmetric resonances. Closed
circles correspond to $L=2a$, open circles correspond to $L=1000a$
and cross corresponds to the monopole Mie resonance 1.}
\label{fig5}
\end{figure}
The $Q$ factor of the antisymmetric monopole resonance
exceeds the $Q$ factor of the isolated cylinder by one order in magnitude.
The reason of that follows from the Mie resonances shown in Fig. \ref{fig3}.
As seen from Fig. \ref{fig5} (a) at the closest distance between the cylinders $L=2a$ the symmetric
resonant mode becomes the monopole mode with corresponding $Q$ factor close to the $Q$ factor
of isolated cylinder which is rather low as marked by cross in Fig. \ref{fig5} (b).
At the same time the antisymmetric mode of two cylinders at
$L=2a$ becomes the dipole resonance which as seen from Fig.
\ref{fig3} has the $Q$ factor  exceeding the $Q$ factor of the
monopole Mie resonance by one order in magnitude.

The next resonances illustrate that their evolution
strongly depend on interaction between the cylinders via the
radiating dipole Mie resonances which are degenerate in the
isolated cylinder. The general expressions and physical origin of
the coupling of  dielectric resonators was considered in Refs.
\cite{Awai2007,Elnaggar2015,Tayebi2018}. The coupling constant can
be written as \cite{Elnaggar2015}
\begin{equation}\label{coupl}
    \kappa=\int dxdy[\epsilon(\overrightarrow{r})-1]\overrightarrow{E}_1^{*}\overrightarrow{E}_2
\end{equation}
where $\overrightarrow{E}_{1,2}$ are normalized solutions by the factor
$\sqrt{\int\epsilon(\overrightarrow{r})|\overrightarrow{E}_{1,2}|^2dxdy}$. Here the indices 1 and 2 imply
the resonant modes of isolated cylinders. One can see that the coupling constant
is determined by overlapping of resonant modes which in turn depend on the distance between the
particles and prevailing direction of radiation of the modes.

\begin{figure}
\includegraphics*[width=8cm,clip=]{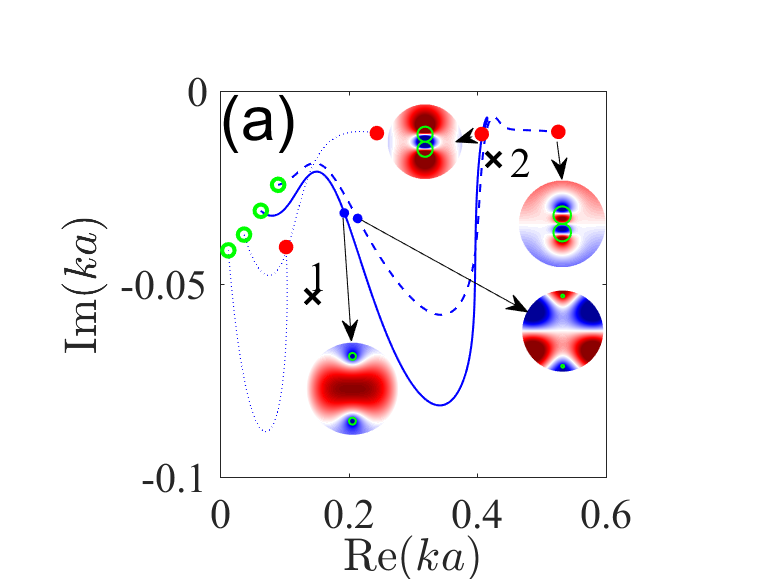}
\includegraphics*[width=8cm,clip=]{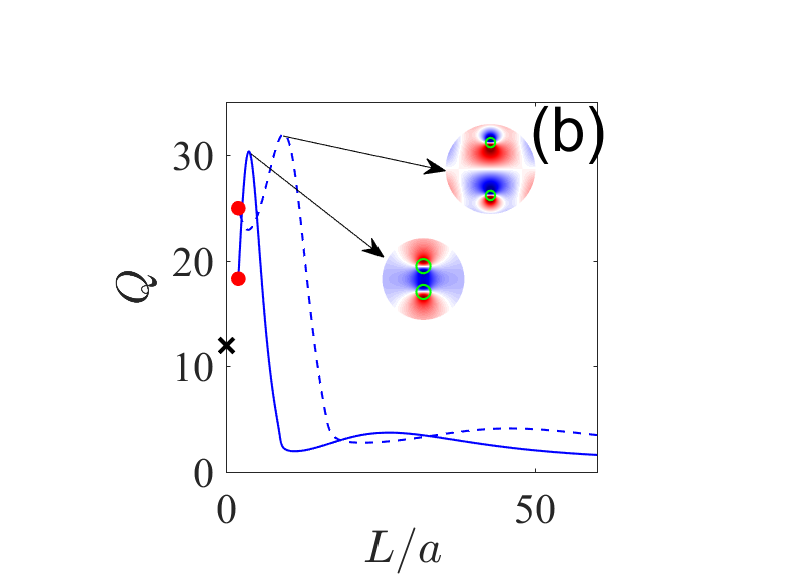}
\caption{The same as in Fig. \ref{fig5} for hybridization of the Mie dipole resonant modes.
Cross marks the $Q$ factor of the isolated cylinder.}
\label{fig6}
\end{figure}

\begin{figure}
\includegraphics*[width=10cm,clip=]{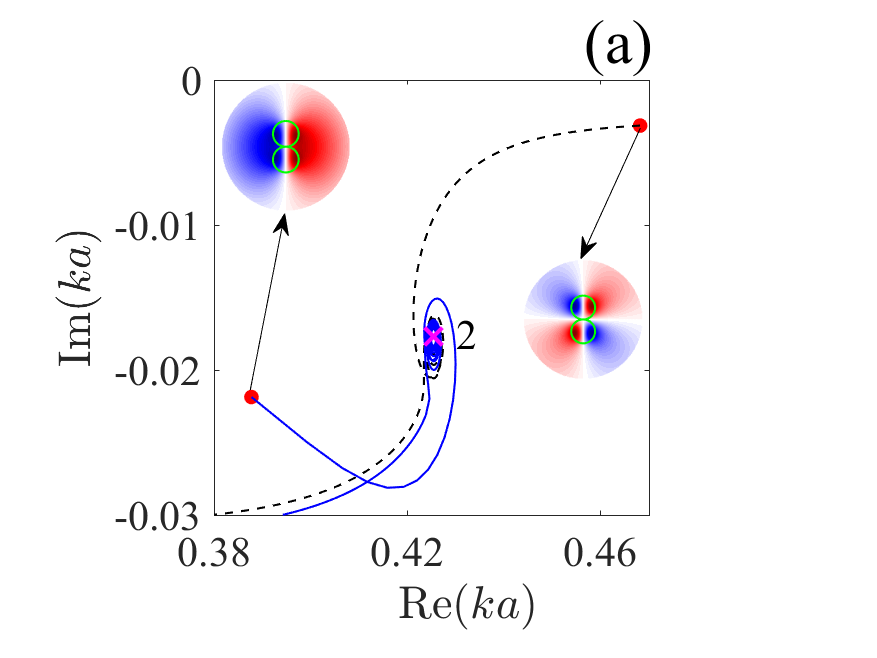}
\includegraphics*[width=9cm,clip=]{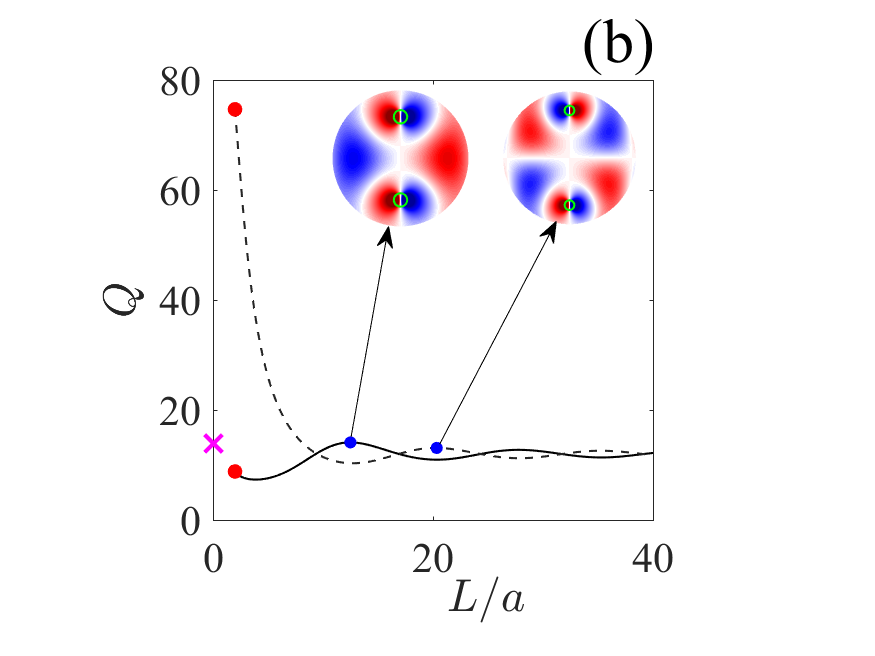}
\caption{The behavior of resonant modes, symmetric and anti symmetric  hybridizations
(\ref{1s,a}) of dipole resonant modes of isolated cylinders (a) and respective $Q$ factors (b)
for variation of distance between them. The dipole Mie resonance is shown by red cross.}
\label{fig7}
\end{figure}

\begin{figure}
\includegraphics[width=8.5cm,clip=]{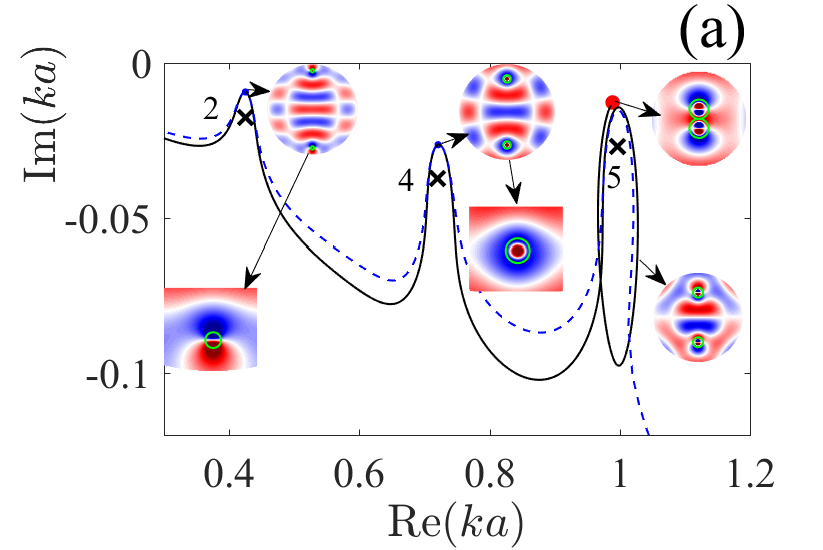}
\includegraphics[width=8cm,clip=]{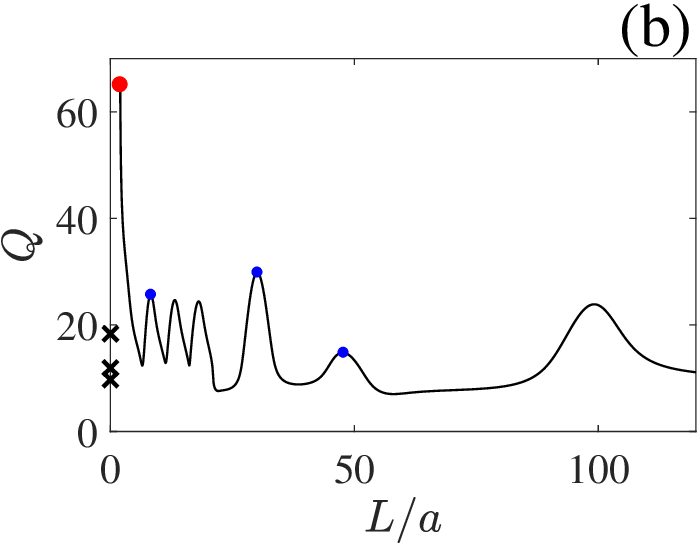}
\caption{The behavior of resonant modes, symmetric and anti symmetric  hybridizations
of dipole resonant modes of isolated cylinders (a) and respective $Q$ factors (b)
for variation of distance between them. The respective dipole Mie resonance is shown by red cross.}
\label{fig8}
\end{figure}

That conclusion is well illustrated by the Mie dipole resonant
modes which are degenerate. The first dipole Mie resonant mode,
symmetric relative to $x\rightarrow -x$, radiates prevalently towards
the neighboring cylinder as shown in insets in Fig. \ref{fig6}
while the second antisymmetric dipole Mie resonant mode
radiates away from the neighboring cylinder as
shown in insets of Fig. \ref{fig7}. As a result the interaction in
former case turns out stronger compared to the latter case as it
follows from Eq. (\ref{coupl}). That explains why the evolution of
resonances shown in Fig. \ref{fig6} (a) is similar to the case of
interaction via the monopole resonant modes in Fig. \ref{fig5}
while the evolution of resonances in Fig. \ref{fig6} is bound to
the Mie dipole resonance 2.
Respectively the gain in the $Q$ factor in the former case is
smaller than in  the latter case as seen from Figs. \ref{fig6} (b)
and \ref{fig7} (b).
With further increase of the distance $L$ the resonance bypasses the
monopole Mie resonance 1 of the isolated cylinder. As a result
the resonant mode becomes close to the
monopole mode $\psi_{s;s/a}$.
In both cases in wide range of distances $L$ except close the resonant modes can be
presented as symmetric and antisymmetric superpositions of the Mie resonant modes
of the isolated cylinder
\begin{eqnarray}\label{1s,a}
    &\psi_{s;s/a}^{(m)}=H_m^{(1)}(kr_1)\cos(m\theta_1)\pm H_m^{(1)}(kr_2)\cos(m\theta_2),&\nonumber\\
    &\psi_{a;s/a}^{(m)}=H_m^{(1)}(kr_1)\sin(m\theta_1)\pm H_m^{(1)}(kr_2)\sin(m\theta_2)&
\end{eqnarray}
except the close distance between the cylinders.
Fig. \ref{fig8} well illustrates Eq. (\ref{1s,a}) for bypassing of resonance
the dipole Mie resonances 2 and 5  with $m=1$ and the monopole Mie resonance 4 $m=0$.

\begin{figure}
\includegraphics*[width=9cm,clip=]{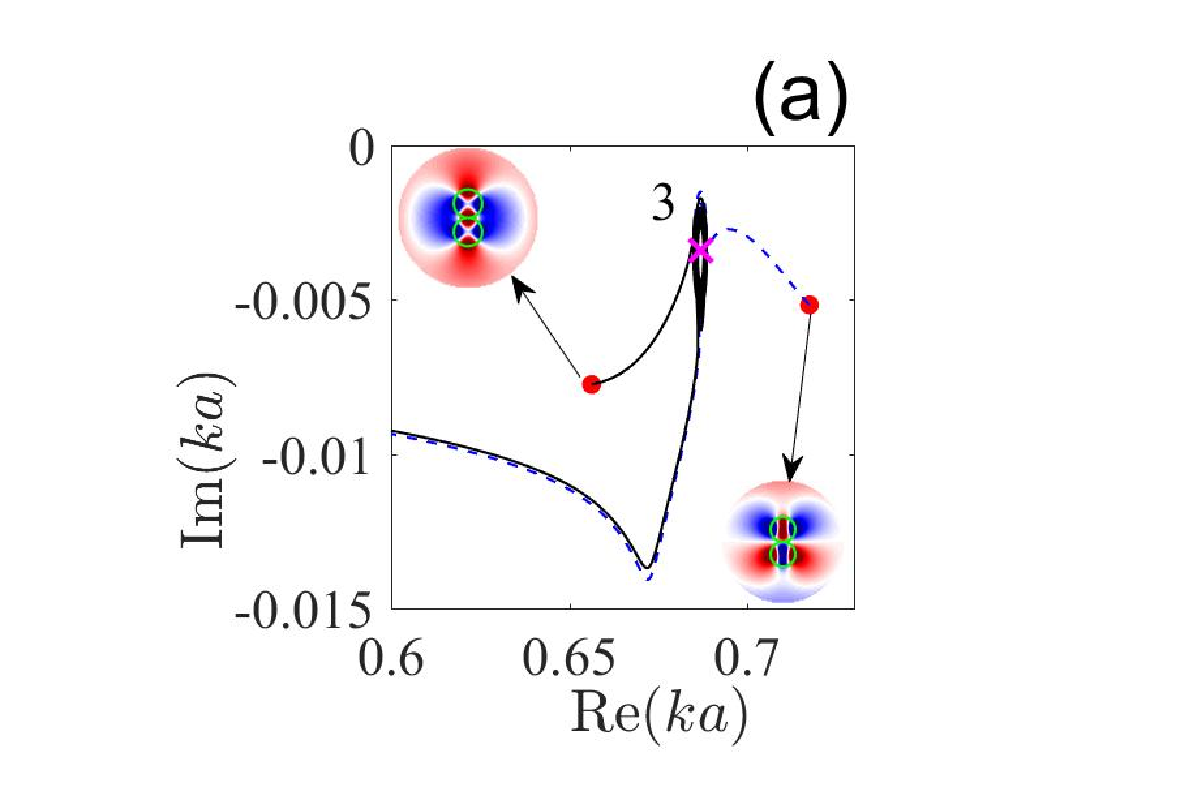}
\includegraphics*[width=9cm,clip=]{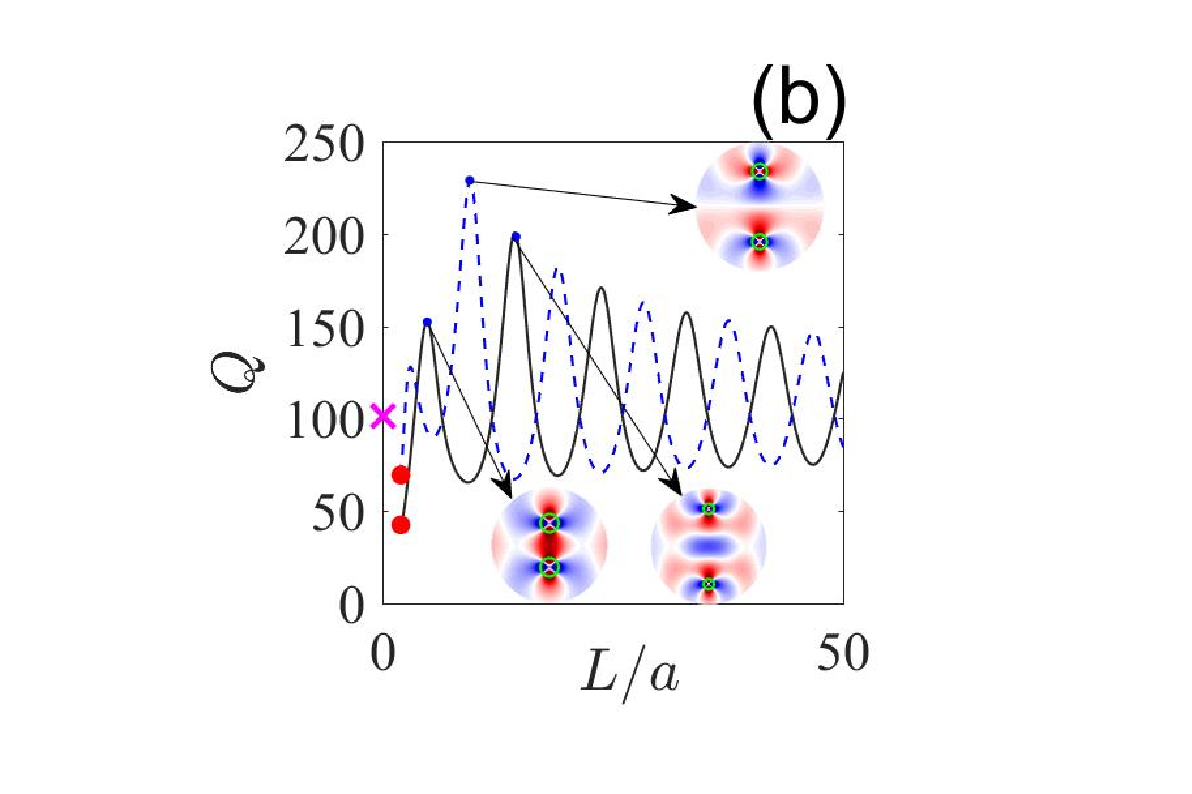}
\includegraphics*[width=9cm,clip=]{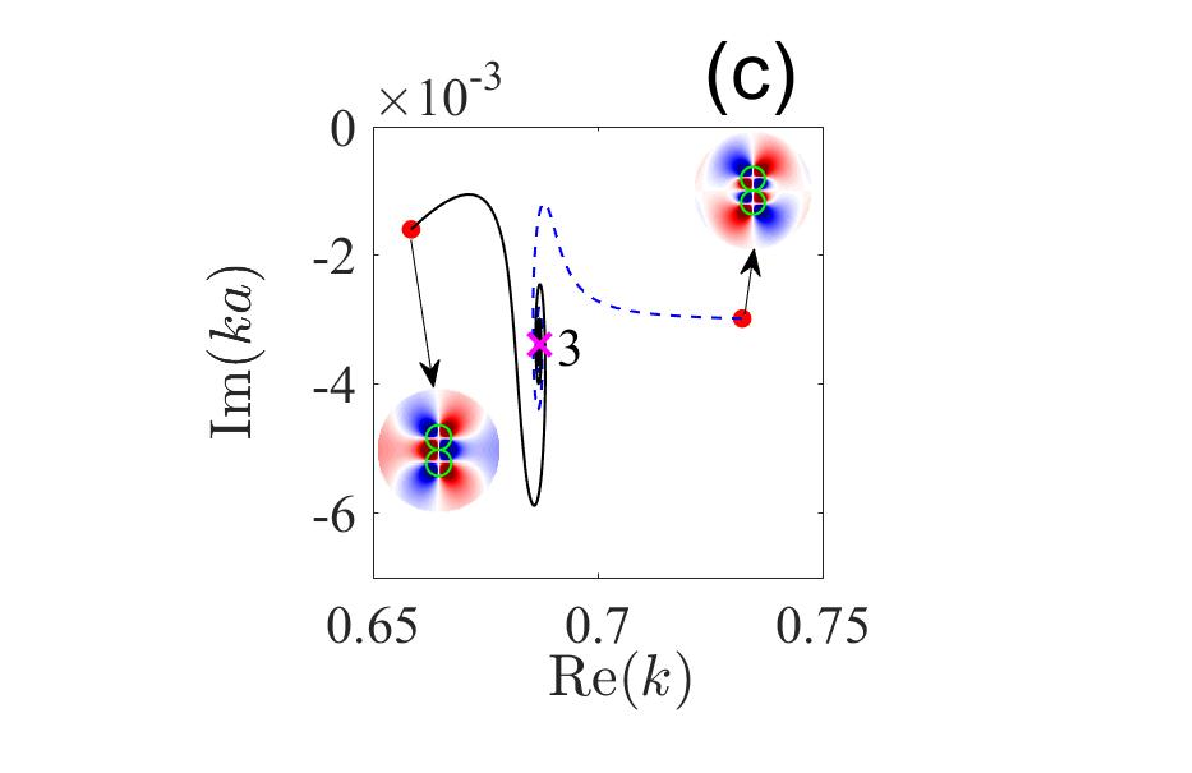}
\includegraphics*[width=9cm,clip=]{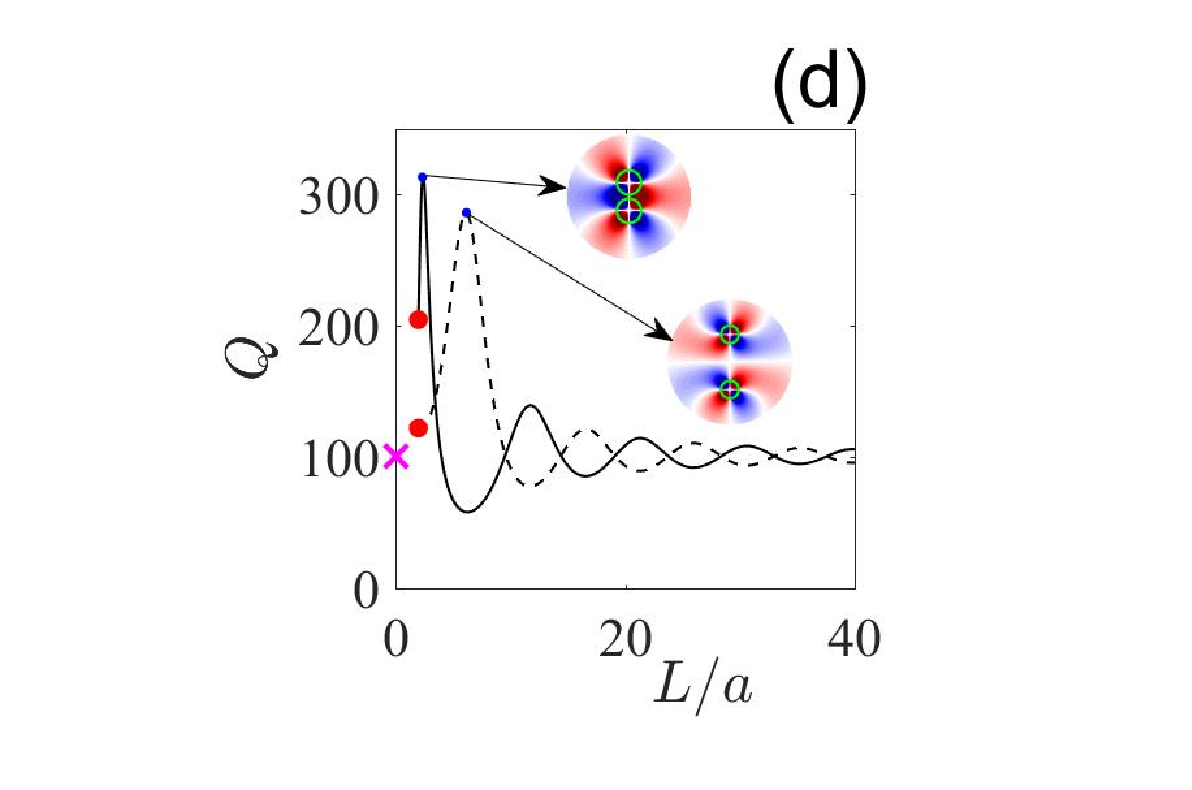}
\caption{The behavior of resonant modes of quadruple resonant modes of isolated cylinders
for variation of distance between them. Crosses mark the Mie resonances
shown in Fig. \ref{fig3}. Red closed circle marks the case of
nearest position of cylinders and open green circle marks the distance $L=1000$.}
\label{fig9}
\end{figure}
\begin{figure}
\includegraphics*[width=8cm,clip=]{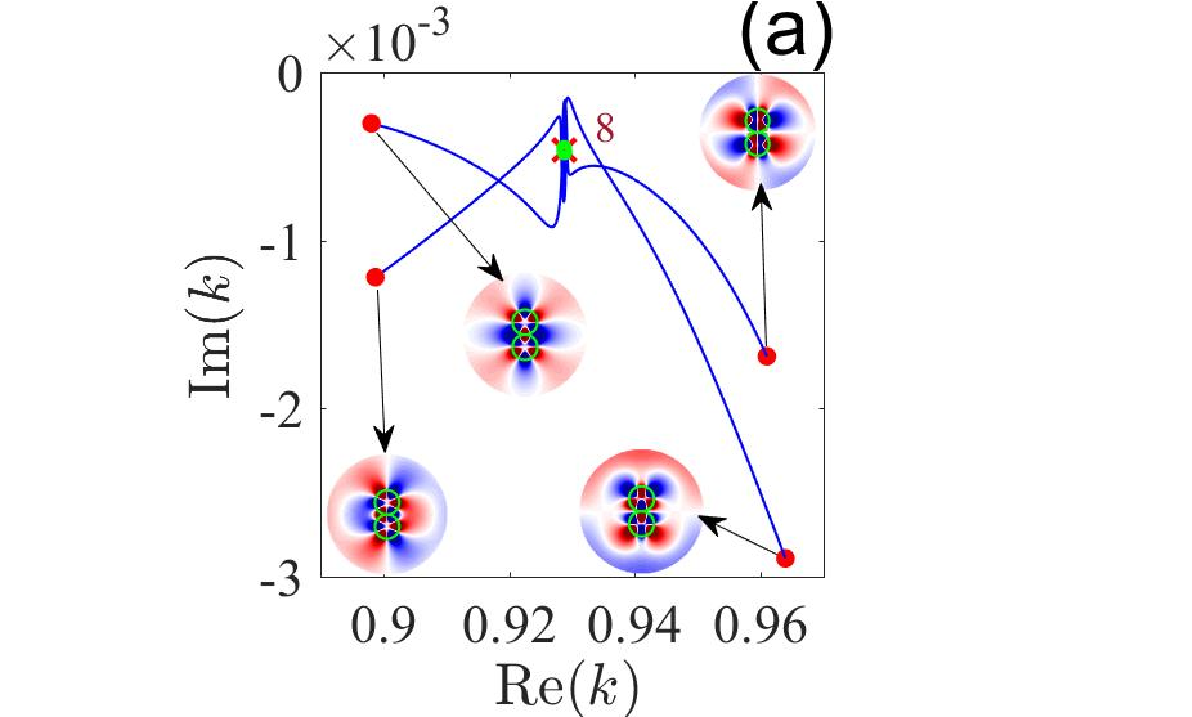}
\includegraphics*[width=8cm,clip=]{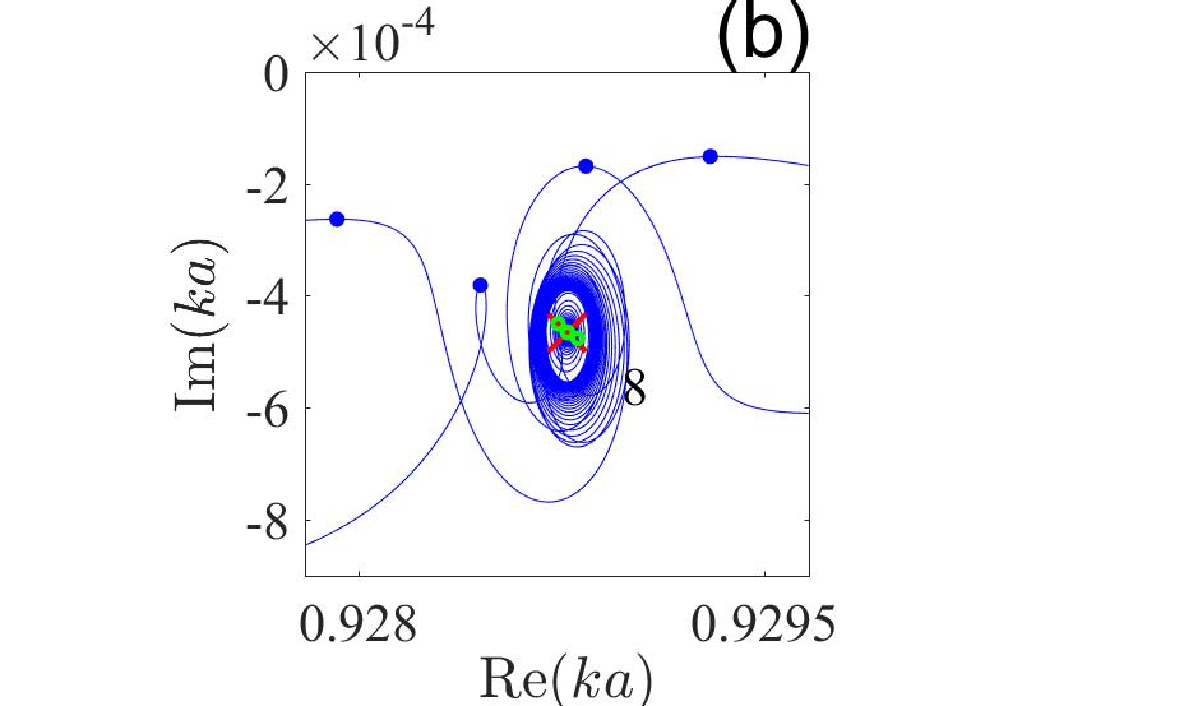}
\includegraphics*[width=8.4cm,clip=]{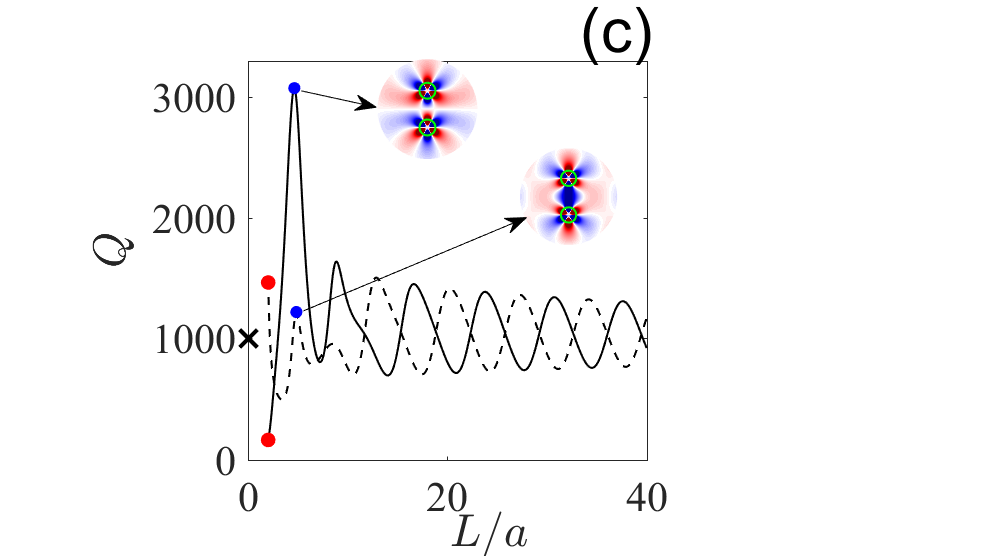}
\includegraphics*[width=8cm,clip=]{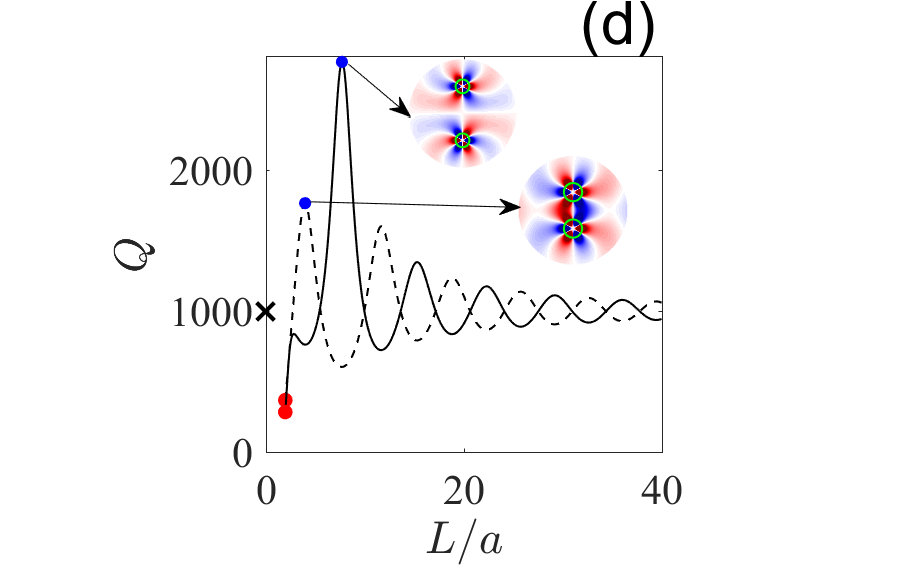}
\caption{The behavior of resonant modes of octuple resonant modes of isolated cylinders
and $Q$ factor for variation of distance between them. Crosses mark the Mie resonances
shown in Fig. \ref{fig3}. Red closed circle marks the case of
nearest position of cylinders and open green circle marks the distance $L=1000$.}
\label{fig10}
\end{figure}

Figs. \ref{fig9} and \ref{fig10} illustrates the second scenario of evolution of resonances which are
bounded by  the  quadruple, octuple {\it etc} Mie  resonances of the isolated cylinder because of
weakness of interaction of cylinders through these resonant modes.
Respectively we observe only oscillating behavior of the $A$ factor with substantial enhancement
compared to the isolated cylinder.
Fig. \ref{fig10} (c), (d) shows as the $Q$ factor can reach extremal enhancement for variation of the
distance similar to the WGM resonances \cite{Wiersig2006,Boriskina2006,Boriskina2007}.

\section{Summary of results and conclusions}
For the isolated dielectric cylinder we have well known Mie resonances specified by
azimuthal index $m=0, \pm 1, \pm 2, \ldots$ (monopole, dipole, quadruple {\it etc} resonances)
due to axial symmetry. Two parallel cylinder have no axial symmetry and therefore the solutions
of homogeneous Maxwell equations are given by series of the Bessel (inside) or Hankel (outside
cylinders) functions in $m$. By the use of Graf formula the coefficients in series satisfy linear algebraic
equations and can be easily found \cite{Olaofe1970,Young1975,Yousif1988,Tsuei1988}.
However there were no studies of behavior of resonances of two cylinders depend on
distance between the cylinders except studies of the $Q$ factor by Boriskina for extremely
highly excited resonances, whispering gallery modes \cite{Boriskina2006,Boriskina2007}.
The study presented in this paper reveals surprisingly complicated behavior of the resonances
with the distance which can be divided into two families. In the first family the resonances
evolve beside the Mie resonances undergoing the avoided crossing. At each event of that
the resonant mode inside the cylinders takes the field profile of the corresponding Mie resonant mode
while the solution between the cylinders takes regular symmetric or antisymmetric packing of
half wavelengths. At these moments the $Q$ factor achieves maximal magnitudes.
This type of evolution of resonances flows is
typical for monopole and those dipole resonance which leakages from one cylinder alongside  the another.

The evolution of resonances bounded by the Mie resonances forms the second family
and typical for higher resonances with $m=2, 3, \ldots$. It is interesting  that the dipole resonance
which leakages apart from the other cylinder unites both families. When the leakage from the
first cylinder is directed to the second the overlapping (\ref{coupl}) that case exceeds the coupling
of the Mie dipole resonant modes which leakage aside the the cylinders. As the result in the first case
the resonances consequently avoid the Mie resonances being unbound while in the second case
the resonances are bound to the dipole Mie resonance of isolated cylinder.
There is a range of distances between
the cylinders where the trajectories of resonances are bounded by the Mie dipole resonances but beyond these distances the flows
are unbounded. For variation of the distance the $Q$ factor shows oscillating behavior with maxima
which can exceed the $Q$ factor of the isolated cylinder three times. That enhancement is typical
for all types of resonances except the monopole resonance which demonstrates enhancement by one order
in magnitude.

{\bf Acknowledgments}: We acknowledge discussions with  D.N.
Maksimov. This  work  was partially supported  by
Ministry  of  Education  and  Science  of  Russian  Federation
(State contract  N  3.1845.2017) and RFBR grant 19-02-00055.

\bibliography{sadreev}
\end{document}